\documentclass[a4paper,twocolumn,superscriptaddress,11pt,accepted=2019-04-21]{quantumarticle}
\pdfoutput=1
\usepackage[utf8]{inputenc}
\usepackage[english]{babel}
\usepackage[T1]{fontenc}
\usepackage{amsmath}
\usepackage{hyperref}
\usepackage{soul}

\usepackage{tikz}
\usepackage{lipsum}

\begin{document}

\title{Quantum process tomography of a high-dimensional quantum communication channel}
\date{\today}
\author{Fr\'ed\'eric Bouchard}
\email{fbouc052@uottawa.ca}
\affiliation{Department of Physics, University of Ottawa, 25 Templeton Street, Ottawa, ON, K1N 6N5, Canada.}
\author{Felix Hufnagel}
\affiliation{Department of Physics, University of Ottawa, 25 Templeton Street, Ottawa, ON, K1N 6N5, Canada.}
\author{Dominik Koutn\'y}
\affiliation{Department of Optics,  Palack\'y  University, 17.~listopadu 12, 771 46 Olomouc, Czech Republic.}
\author{Aazad Abbas}
\affiliation{Department of Physics, University of Ottawa, 25 Templeton Street, Ottawa, ON, K1N 6N5, Canada.}
\author{Alicia Sit}
\affiliation{Department of Physics, University of Ottawa, 25 Templeton Street, Ottawa, ON, K1N 6N5, Canada.}
\author{Khabat Heshami}
\affiliation{National Research Council of Canada, 100 Sussex Drive, Ottawa, Ontario K1A 0R6, Canada}
\author{Robert Fickler}
\affiliation{Department of Physics, University of Ottawa, 25 Templeton Street, Ottawa, ON, K1N 6N5, Canada.}
\affiliation{Institute for Quantum Optics and Quantum Information (IQOQI), Austrian Academy of Sciences, Boltzmanngasse 3, A-1090 Vienna, Austria.}
\affiliation{Current address: Photonics Laboratory, Physics Unit, Tampere University, Tampere, FI-33720, Finland.}
\author{Ebrahim Karimi}
\affiliation{Department of Physics, University of Ottawa, 25 Templeton Street, Ottawa, ON, K1N 6N5, Canada.}
\affiliation{Department of Physics, Institute for Advanced Studies in Basic Sciences, 45137-66731 Zanjan, Iran.}

\begin{abstract}
 \begin{abstract}
The characterization of quantum processes, e.g. communication channels, is an essential ingredient for establishing quantum information systems. For quantum key distribution protocols, the amount of overall noise in the channel determines the rate at which secret bits are distributed between authorized partners. In particular, tomographic protocols allow for the full reconstruction, and thus characterization, of the channel. Here, we perform quantum process tomography of high-dimensional quantum communication channels with dimensions ranging from 2 to 5. We can thus explicitly demonstrate the effect of an eavesdropper performing an optimal cloning attack or an intercept-resend attack during a quantum cryptographic protocol. Moreover, our study shows that quantum process tomography enables a more detailed understanding of the channel conditions compared to a coarse-grained measure, such as quantum bit error rates. This full characterization technique allows us to optimize the performance of quantum key distribution under asymmetric experimental conditions, which is particularly useful when considering high-dimensional encoding schemes.
\end{abstract}
\end{abstract}

\section{Introduction}
Quantum information science has witnessed the emergence of a wide range of new technologies and applications~\cite{dowling:03}. Quantum cryptography~\cite{bennett:84}, quantum computation~\cite{nielsen:02} and quantum sensing~\cite{degen:17} are examples of promising venues for a possible next technological revolution. In order to construct complex quantum machines or quantum networks, a full characterization of its building blocks is critical. A method for reconstructing the action of a component in a quantum system is known as \emph{quantum process tomography} (QPT)~\cite{chuang:97,poyatos:97}. Previously, QPT has been performed to characterize several quantum physical systems, such as liquid-state NMR~\cite{nielsen:98,childs:01}, photonic qubits~\cite{obrien:04}, atoms in optical lattices~\cite{myrskog:05}, trapped ions~\cite{riebe:06}, solid-state qubits~\cite{howard:06}, continuous-variable quantum states~\cite{lobino:08}, semiconductor quantum dot qubits~\cite{kim:14} and, recently, nonlinear optical systems~\cite{jacob:18}. Another class of important quantum systems that can benefit from full characterization are quantum channels and components for quantum key distribution (QKD) and quantum communications~\cite{gisin:02,ndagano:17}, where, so far, quantum channels may be categorized as optical fibre~\cite{muller:95}, line-of-sight free-space~\cite{buttler:98} and ground-to-satellite (satellite-to-ground)~\cite{yin:17} links. 

The benefits of fully characterizing quantum communication channels lie at a better understanding of possible error sources and, more importantly, on the detection of the potential presence of an eavesdropper, namely \emph{Eve}, tapping into the quantum channel. Her presence is revealed in the form of noise introduced in the channel. The authorized partners, typically referred to as \emph{Alice} and \emph{Bob}, may estimate noise levels in the channel to assess the amount of leaked information to Eve. After this error assessment, Alice and Bob may perform classical post-processing protocols, such as \emph{privacy amplification}~\cite{bennett:95}, in order to remove Eve's leaked information. Conventional methods for secure key rate analysis rely on symmetry assumptions to associate an average (coarse-grained over all preparation and measurement settings) bit error rate parameter to a reduction in secure key rates. However, experimental errors often tend to break these symmetry assumptions, motivating the use of numerical techniques in QKD~\cite{coles:16}. 

While usual QKD schemes rely on encoding quantum information in two-level systems, i.e. qubits, there exists a particular class of QKD schemes known as \emph{high-dimensional} QKD protocols~\cite{bechmann:00b,cerf:02}. High-dimensional schemes have the potential advantage of tolerating larger noise levels in the channel and carrying more than one bit of information per carrier. So far, a full characterization of high-dimensional processes based on QPT has not been achieved experimentally. With the emergence of high-dimensional quantum information, QPT will become an essential tool for the characterization of complex experiments dealing with high-dimensional quantum states. As a physical implementation, the orbital angular momentum (OAM) of photons represents a promising route for high-dimensional encoding due to the maturity of its generation and detection schemes~\cite{molina:07,erhard:17}. OAM states, corresponding to helical wavefronts of the form $\exp (i \ell \varphi)$, where the OAM value $\ell$ is an integer and $\varphi$ is the azimuthal coordinate, may be realized using single phase elements. Computer-generated holograms displayed on a spatial light modulator (SLM) provide a simple and versatile method for generating and manipulating these modes~\cite{forbes:16}. High-dimensional QKD has been demonstrated experimentally using OAM in the laboratory~\cite{mafu:13,mirhosseini:15,bouchard:18b}, in intra-city free-space links~\cite{vallone:14,krenn:15,sit:17} and, recently, in other types of quantum links~\cite{bouchard:18}.

Here, we demonstrate the benefits of a full characterization of a quantum channel through quantum process tomography. This full characterization will allow us to complement the numerical approach in~\cite{coles:16} to optimize secure key rates under specific experimental conditions and to develop new protocols lacking symmetry that may outperform existing approaches. 

\begin{figure}[ht!]
	\begin{center}
	\includegraphics[width=1.0\columnwidth]{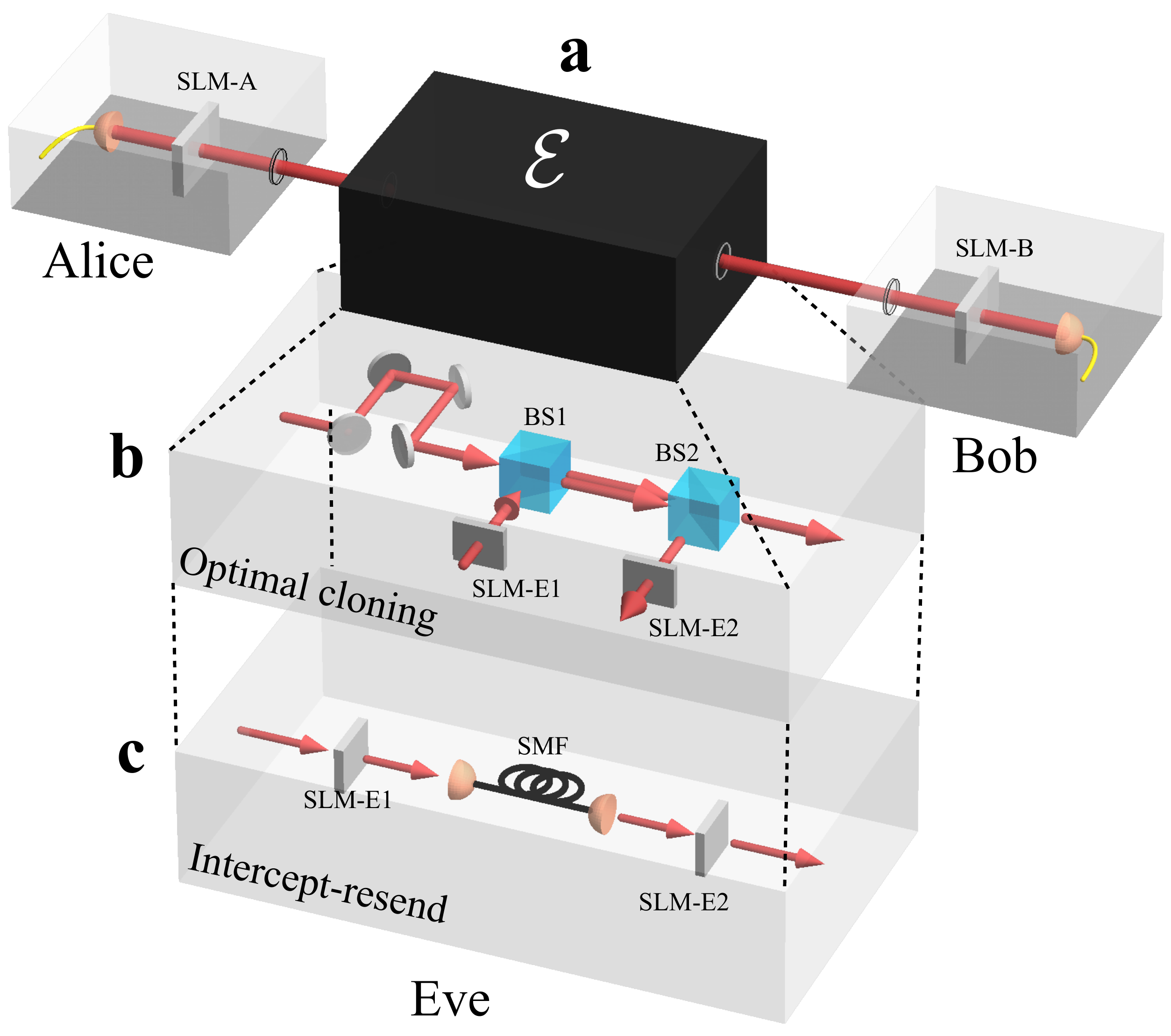}
	\caption[]{\textbf{Simplified experimental setup.} \textbf{a} Alice's preparation stage consists of a single photon source (not shown) that feeds the heralded photons to a generation apparatus consisting of a spatial light modulator (SLM-A). Alice's photon is subsequently sent into the quantum channel which is considered as a black box with quantum process ${\cal E}$. The output of the black box is fed to Bob's detection stage consisting of a spatial light modulator (SLM-B) and a single-mode optical fibre (SMF), via performing phase-flattening. We experimentally consider three different types of processes for our quantum channel: an ideal channel with no attack, the case of an optimal cloning attack (\textbf{b}), and the case of an intercept-resend attack (\textbf{c}). In particular, the configuration of the optimal cloning scheme consists of an SLM (SLM-E1) which generates a completely mixed state that is fed to the Hong-Ou-Mandel type interferometer at a balanced beam splitter (BS1). The \emph{cloned} photons are then sent into a second beam splitter (BS2) in order to spatially separate them. Eve can use a second SLM (SLM-E2) to measure the state of her cloned photon while sending the other cloned photon to Bob. The experimental configuration of the intercept resend consists of two SLMs (SLM-E1 and SLM-E2) sandwiching a SMF.}
	\label{fig:setup}
	\end{center}
\end{figure}

\begin{figure*}[ht!]
	\begin{center}
	\includegraphics[width=1.5\columnwidth]{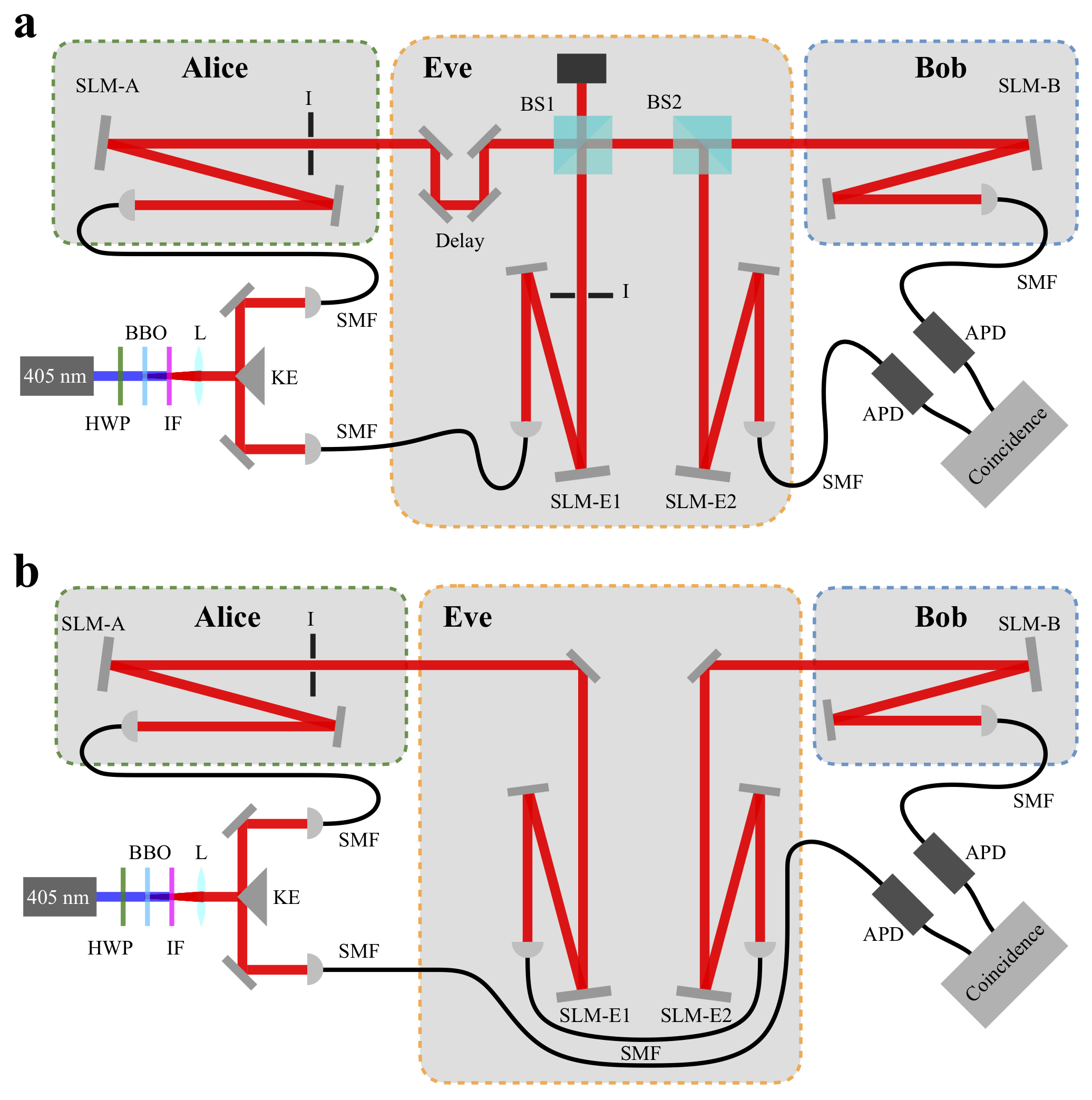}
	\caption[]{\textbf{Detailed experimental setup for (a) the optimal cloning attack and (b) the intercept-resend attack}. Single photons are generated, via spontaneous parametric downconversion (SPDC), by pumping a BBO crystal with a 355~nm quasi-continuous laser. The polarization of the laser is chosen in such a way to satisfy the phase-matching condition of the crystal. Subsequent to SPDC, the 355~nm laser is filtered out using an interference filter (IF). By using a lens (L), the photon pairs are split at the knife-edge (KE) mirror. The photons are then made to couple to single mode fibres (SMF) to filter their spatial modes to the fundamental Gaussian mode. Alice encodes the photon by using the spatial light modulator (SLM-A), and sends it to the quantum channel. Eve may perform an attack by using a delay line, beam splitters (BS1 and BS2), spatial light modulators (SLM-E1 and SLM-E2), and SMF. Bob measures the photon coming from the quantum channel by using his spatial light modulator (SLM-B). Photons counts are recorded at the avalanche photodiodes (APD) detects and coincidence measurements are recorded using a coincidence box.}
	\label{fig:det_exp_setup}
	\end{center}
\end{figure*}

\section{Theory of quantum process tomography}
Let us set the stage by introducing the basic concepts of QPT. The goal of QPT is to determine the completely positive map ${\cal E}$ representing the action of the system on any $d$-dimensional input state $\rho_\mathrm{in}$. The output state is then given by $\rho_\mathrm{out} = {\cal E} \left( \rho_\mathrm{in} \right)$. The process is typically represented as,
\begin{equation}
	{\cal E} (\rho) = \sum_{m,n} \chi_{mn} \, \hat{A}_m \, \rho \, \hat{A}^\dagger_n,
\end{equation}
where the $\hat{A}_m$ operators form a complete basis typically given by the Pauli operators or the Gell-Mann operators in higher dimensions. The trace-preserving positive Hermitian $d^2 \times d^2$ matrix $\chi_{mn}$, defined as the \emph{process matrix}, completely and uniquely characterizes the action of the process ${\cal E}$. We may note that QPT may also be performed for non-trace-preserving maps~\cite{bongioanni:10}.

A convenient alternative description of processes is given by the \emph{Choi-Jamiolokowski} isomorphism (CJI), which states that every completely positive map can be represented as an operator living in a $d^2$-dimensional Hilbert space. Such an operator, known as the Choi matrix $\rho_{\cal E}$, can be defined as the result of the channel acting upon one part of a maximally entangled state, $|\Phi \rangle$,
\begin{eqnarray}
	\rho_{\cal E} = \left( \hat{I}\, \otimes \, {\cal E} \right) | \Phi \rangle \langle \Phi |.
	\label{eq:choi}
\end{eqnarray}
The output state is thus given accordingly by,
\begin{equation}
\rho_{\mathrm{out}} = \mathrm{Tr}_{\mathrm{in}} \left[ \, \left( \rho_{\mathrm{in}}^T\otimes \hat{I} \right) \, \rho_{\cal E} \right],
\end{equation}
where $T$ denotes the transposition, $\hat{I}$ is the $d$-dimensional-identity operator and $\mathrm{Tr}_{\mathrm{in}} \left[\cdot \right]$ represents the partial trace over the input state's Hilbert space. In order to perform QPT, a set of tomographically complete states are sent into the channel and state tomography is performed on the output states. In our case, we consider states belonging to mutually unbiased bases (MUBs) in dimension $d$, which are known for dimensions that are powers of prime numbers~\cite{wootters:89}. A MUB projector is given by $\Pi^{(\alpha)}_m$, {where $\alpha \in \{1,...,d+1 \}$ labels one of the MUBs and $m \in \{ 1,...,d\}$ labels one of the states in this MUB}. The completeness and orthogonality relations are respectively given by, $ \sum_{\alpha,m} \Pi^{(\alpha)}_m /(d+1) = \hat{I}$ and $\mathrm{Tr} (\Pi^{(\alpha)}_m \Pi^{(\beta)}_n) = \delta_{\alpha \beta} \delta_{mn}+(1-\delta_{\alpha \beta})/d$, where $\delta_{ij}$ is the Kronecker delta. With the help of the CJI, one can derive Born rule--like expressions for probabilities,
\begin{eqnarray}
p^{(\alpha, \,\beta)}_{m,n} &=& \mathrm{Tr}\left[ \Pi^{(\beta)}_n {\cal E} \left( \Pi^{(\alpha)}_m \right) \right]\nonumber \\ &=& \mathrm{Tr}\left[ \, \left( \left( \Pi^{(\alpha)}_m \right)^T \otimes \Pi^{(\beta)}_n \right) \rho_{\cal E} \, \right].
\label{probability2}
\end{eqnarray}
This equation underlies the relation between quantum process tomography and quantum state tomography. The Choi matrix, $\rho_{\cal E}$, is experimentally reconstructed using the maximum--likelihood estimation. Better algorithms have also been recently proposed for QPT~\cite{knee:18}. Finally, the process matrix is directly obtained from the Choi matrix.\newline

{In order to recover the Choi matrix, $\rho_{\cal E}$, defined in Eq.~\ref{eq:choi}, for various attack scenarios, we follow maximum likelihood algorithms. The positive operator-valued measure (POVM) used to describe the process matrix consists of MUB measurements, i.e. $\Pi_j = (\Pi^{(\alpha)}_m)^T \otimes \Pi^{(\beta)}_n$, where the index $j$ runs over all the MUBs measurements performed by Alice and Bob. Theoretical probabilities given by the Born rule, $p_j = \mathrm{Tr}[\Pi_j \rho_{\cal E}]$, observed with relative frequencies $f_j$, can be inverted by maximizing the log--likelihood functional
\begin{equation}\label{eq4}
\log \mathcal{L} = \sum_j f_j \log \{p_j(\hat{\rho}_{\cal E})\},
\end{equation}
yielding the Choi matrices. Any density operator maximizing the likelihood functional satisfies the extremal equation ${\hat{R} \, \hat{\rho}_{\cal E} = \hat{\rho}_{\cal E}}$, which leads to the well known iterative solutions~\cite{jezek:03} that takes a symmetric form,
\begin{equation}\label{eq:map}
\hat{\rho}_{\cal E}^{(k+1)} = \mu \, \hat{R} \, \hat{\rho}_{\cal E}^{(k)} \, \hat{R},
\end{equation}
where $\mu = \mathrm{Tr}[\hat{R} \, \hat{\rho}_{\cal E}^{(k)} \, \hat{R}]^{-1}$ is a proper normalization constant and the operator $\hat{R}$ is defined as follows,
\begin{equation}\label{eq6}
\hat{R}=\sum\limits_j\frac{f_j}{p_j(\hat{\rho}_{\cal E})}\hat{\Pi}_j.
\end{equation}
Typically, a few thousands of iterations are needed to observe the stationary point of the map in Eq.~\ref{eq:map} with the maximally mixed state as a suitable choice of a starting point.}

\begin{figure*}[ht!]
	\begin{center}
	\includegraphics[width=1.8\columnwidth]{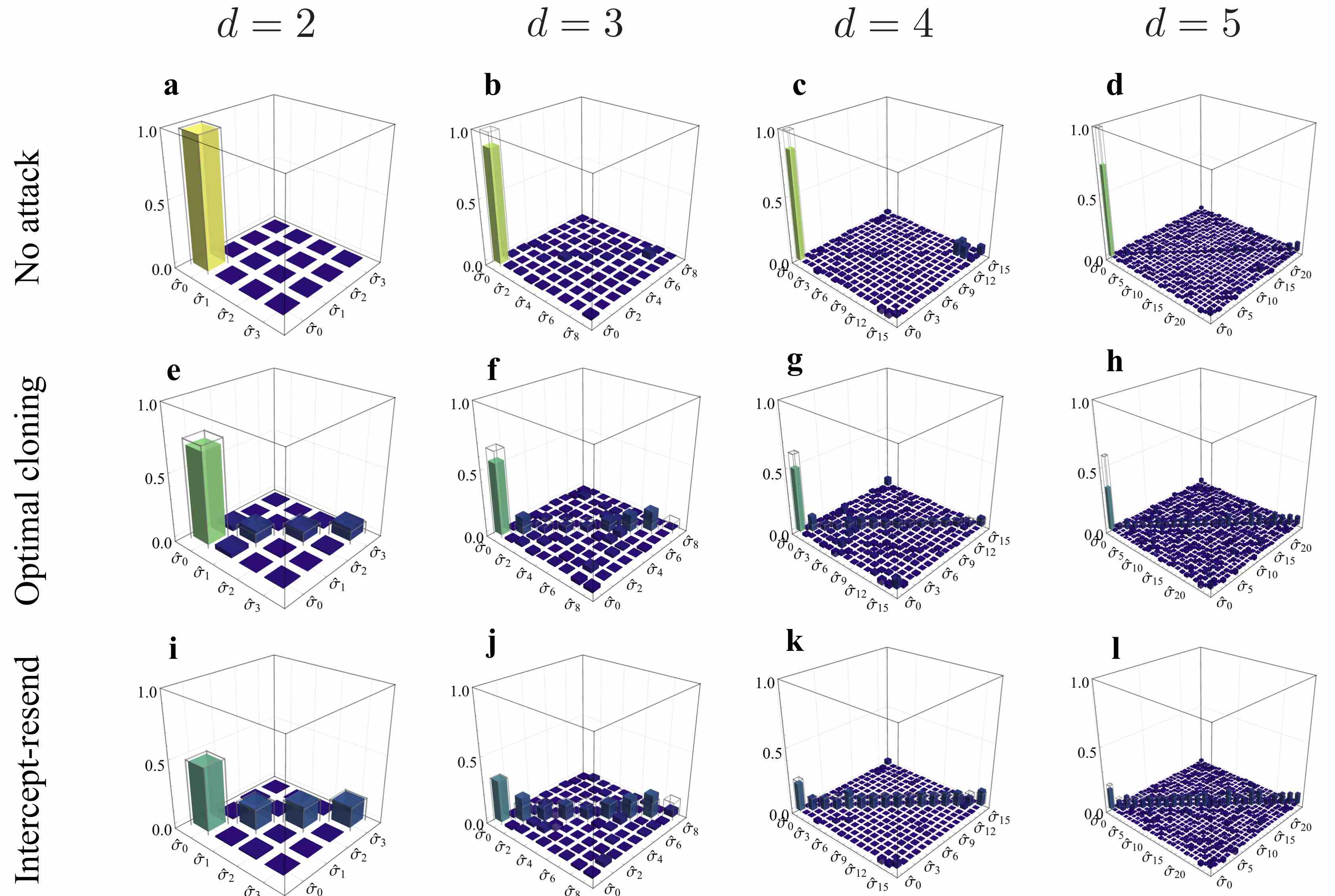}
	\caption[]{\textbf{Process matrix of high-dimensional quantum channels}. The real part of the experimentally reconstructed process matrices are presented for dimensions ranging from 2 to 5 in the case of (\textbf{a}-\textbf{d})  an ideal quantum channel, (\textbf{e}-\textbf{h}) a universal symmetric optimal quantum cloning attack, and (\textbf{i}-\textbf{l}) a universal intercept-resend attack. 
	The theoretical results are represented by the wire-grid bars.}
	\label{fig:chi}
	\end{center}
\end{figure*}

\section{Experimental results}

Our experiment consists of three components: the generation stage, the quantum channel and the detection stage, which are owned by Alice, Eve and Bob, respectively. We implement the prepare-and-measure QKD scheme with heralded single-photons using the OAM degree of freedom, see Fig.~\ref{fig:setup}-\textbf{a}. The single photon pairs, consisting of the \emph{signal} and \emph{idler}, are generated by spontaneous parametric downconversion (SPDC) at a type I $\beta$-barium borate (BBO) crystal. The nonlinear crystal is pumped by a quasi-continuous wave laser operating at a wavelength of 355~nm. The generated photon pairs are coupled to single-mode optical fibres (SMF) in order to filter their spatial modes to the fundamental Gaussian mode. A coincidence rate of 30~kHz is measured, directly at the source, within a coincidence time window of 2~ns. We use the detection of the idler photon as the heralding trigger for the signal photon, thereby realizing a heralded single photon source with a measured second-order coherence of $g^{(2)}(0) = 0.015 \pm 0.004$. The heralded signal photon is sent to SLM-A corresponding to Alice's generation stage. The OAM states are produced using a phase-only holography technique~\cite{forbes:16}. Alice's heralded photon is subsequently sent over the quantum channel, considered here as a black box between Alice and Bob. At Bob's receiver, we perform a state-projection over the required states, which we realize by a mode filter implemented through a phase-flattening hologram and coupling into a SMF~\cite{mair:01,qassim:14}. {In order to minimize Poissonian noise, the counts are accumulated over an integration time of 60 seconds per measurement setting. A detailed experimental setup is shown in Fig.~\ref{fig:det_exp_setup}.} In particular, Alice and Bob perform QPT using MUBs~\cite{fernandez:11}, which are experimentally realized using OAM states. The computational bases ($\alpha=1$) are given by $\{ | \ell \rangle ; \ell = -d/2, ... , d/2 ; \ell \neq 0 \}$ and $\{ | \ell \rangle ; \ell = -(d-1)/2, ... , (d-1)/2 \}$ for even and odd dimensions, respectively, for symmetry considerations. The second basis is given by the discrete Fourier transform, i.e. $\{| \phi_k \rangle = \frac{1}{\sqrt{d}} \sum_{j=0}^{d-1} \omega_d^{kj} |j\rangle ; k=0,...,d-1  \}$, where $\omega_d = \exp (i 2 \pi /d)$. The explicit form of the other MUB elements can be found elsewhere~\cite{durt:10}.

\begin{table*}
\centering
	\begin{tabular*}{\textwidth}{c @{\extracolsep{\fill}} ccccc}
	\hline \hline
	Attack & Dimension - $d$ & Process fidelity - $F_p$ & Average state fidelity - $\bar{F}$ & Process purity - $\bar{P}$\\
	\hline 
	No attack & 2 & $0.990 \pm 0.001$ & $0.993 \pm 0.001$ & $0.987 \pm 0.001$  \\
	& 3 & $0.907 \pm 0.003$ & $0.930 \pm 0.002$ & $0.868 \pm 0.004$ \\
	& 4 & $0.888 \pm 0.003$ & $0.911 \pm 0.002$ & $0.832 \pm 0.004$ \\
	& 5 & $0.821 \pm 0.002$ & $0.851 \pm 0.002$ & $0.729 \pm 0.003$ \\
	\hline
	Optimal cloning & 2 & $0.71 \pm 0.01$ & $0.806 \pm 0.006$ & $0.687 \pm 0.007$  \\
	& 3 & $0.604 \pm 0.005$ & $0.703 \pm 0.003$ & $0.538 \pm 0.004$ \\
	& 4 & $0.524 \pm 0.007$ & $0.619 \pm 0.006$ & $0.432 \pm 0.006$ \\
	& 5 & $0.419 \pm 0.003$ & $0.511 \pm 0.003$ & $0.321 \pm 0.002$ \\
	\hline
	Intercept-resend & 2 & $0.488 \pm 0.006$ & $0.659 \pm 0.004$ & $0.550 \pm 0.003$  \\
	& 3 & $0.334 \pm 0.002$ & $0.500 \pm 0.001$ & $0.375 \pm 0.001$ \\
	& 4 & $0.226 \pm 0.002$ & $0.381 \pm 0.002$ & $0.273 \pm 0.001$ \\
	& 5 & $0.171 \pm 0.001$ & $0.309 \pm 0.001$ & $0.215 \pm 0.001$ \\
	 \hline \hline
	\end{tabular*}
\caption[]{\textbf{Quality of the quantum channel processes.} Process fidelity, average fidelity and average purity of the quantum channel under different eavesdropping strategies, i.e. optimal cloning attack and intercept-resend attack, for dimensions ranging from 2 to 5.}
\label{table:1}
\end{table*}

\subsection{Ideal quantum communication channel} 
In the case of an ideal quantum channel, i.e. no eavesdropper, the real parts of the experimentally reconstructed process matrices are shown in Fig.~\ref{fig:chi}-(\textbf{a-d}) for dimensions ranging from $d=2$ to 5. Ideal processes, where $\tilde{{\cal E}} (\rho) = \rho$, are described by process matrices given by $\tilde{\chi}_{ij}=\delta_{0i}\delta_{0j}$. Deviations of our reconstructed ideal process matrices from the theory are attributed to imperfections in generation and detection of the distributed quantum states, i.e. the OAM modes. As higher dimensions are considered, larger crosstalk among the OAM modes is observed, which further deteriorates the experimentally reconstructed process matrix. Recently, we have introduced a new technique to measure spatial modes, which should improve detection fidelities~\cite{bouchard:18c}. In order to describe the quality or performance of the quantum channel, several figures of merit are available to describe our process~\cite{gilchrist:05}. One such figure is the \emph{process fidelity} which is defined as $F_P = \mathrm{Tr} (\chi_\mathrm{exp} \, \tilde{\chi})$. For the ideal quantum channel, $F_P$ can be directly obtained from the process matrix $\chi_\mathrm{exp}$, i.e. $F_P=\chi_{i=0,j=0}$. Moreover, the \emph{average fidelity}, $\bar{F}$, of the process can be defined as the state fidelity between the output and the input averaged over all possible states with the convenient relation $\bar{F} = (d F_P +1)/(d+1)$. Similarly, the \emph{average purity} is defined as the purity of the outgoing states averaged over all possible states and describes the level of mixture introduced by the process. It can be related to the average fidelity according to $\bar{P}=(1-2 \bar{F}+d \bar{F}^2)/(d-1)$, respectively. The experimentally obtained process fidelities, average fidelities and average purities for the ideal channel are shown in Table~\ref{table:1} for dimensions ranging from $d=2$ to 5.

\subsection{Optimal quantum cloning attack} 
A special case of such a process is the so-called universal optimal cloning attack. In this scenario, Eve sends the incoming photon to an optimal cloning machine which produces two imperfect copies of the incoming photon, out of which she keeps one and sends the other to Bob, see Fig.~\ref{fig:setup}-\textbf{b}. Later on, Eve may decide to perform a measurement on her copy in order to obtain information from Alice and Bob's shared photon. For the case of universal optimal cloning, Eve's copying machine has the effect of symmetrically introducing errors on the outgoing state. The cloning fidelity is defined as $F_\mathrm{cl}=1/2+1/(1+d)$. Universal optimal cloning machines have been realized experimentally using the symmetrization method~\cite{irvine:04,ricci:04,nagali:09,nagali:10,bouchard:17}, where both cloned photons possess the same cloning fidelity. At the heart of the symmetrization technique for optimal cloning is the Hong-Ou-Mandel interference effect~\cite{hong:87}. For indistinguishable photons incident at the input port of a balanced beam splitter, a two-photon interference effect occurs which forbids the photons to exit the beam splitter from different output ports. Surprisingly, a slight modification of the Hong-Ou-Mandel experiment leads to a universal optimal symmetric cloning machine. In this case, the second photon needs to be in a $d$-dimensional completely mixed state and is fed into the second input port of the beam splitter at exactly the same time. As the representation of the mixed state is the same in any basis, this scheme is universal, i.e. works for any state of the Hilbert space. When the two photons exit the beam splitter from the same output port, optimal cloning has been successful. A second beam splitter is then used to separate the two photons, e.g. one photon goes to Bob and the other one stays at Eve. We then experimentally perform the universal optimal cloning machine in the quantum channel for dimensions up to 5 and perform, again, a full characterization of the attack on the channel through QPT, see Fig.~\ref{fig:chi}-(\textbf{e-h}) and Table~\ref{table:1}. {Leaving attacks on QKD systems aside, a full characterization of optimal quantum cloning machines is interesting in itself for applications to quantum state estimation~\cite{bruss:99,guillaume:17}. Indeed, there is an equivalence between a universal optimal cloning machine taking $N$ input replicas and outputting an infinite number of clones and a quantum state estimation setup. A full characterization of the cloning machine would then enable a higher mode purity in the state estimation task.} \newline

\begin{figure*}[ht!]
	\begin{center}
	\includegraphics[width=1.9\columnwidth]{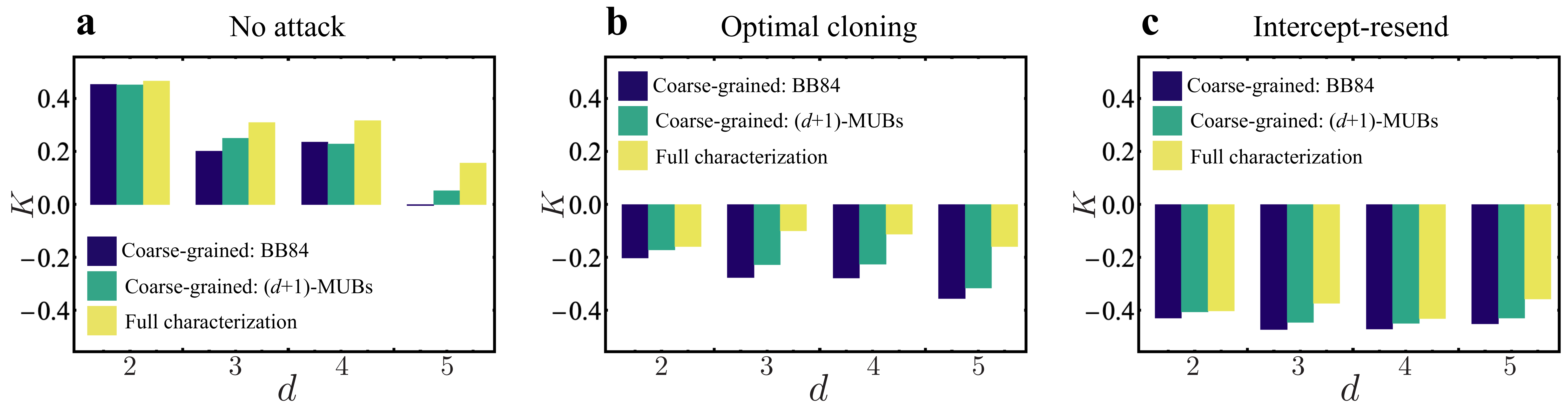}
	\caption[]{{\textbf{Secret key rates.} Experimentally obtained secret key rates for the cases of no attack, optimal cloning and intercept-resend. The dark blue bars correspond to the BB84 protocol where the secret key rate is obtained by estimating a coarse-grained error rate. The green bars correspond to the $(d+1)$-MUBs protocol where the secret key rate is obtained also by estimating a coarse-grained error rate. The yellow bars correspond to a secret key rate calculated from the full characterization of the quantum channel. In all cases, a filter type measurement is considered yielding an overall efficiency of $1/d$.}}
	\label{fig:keyrate}
	\end{center}
\end{figure*}

\subsection{Intercept-resend attack} 
Another eavesdropping scheme that achieves processes with lower average fidelities is the so-called intercept-resend attack, see Fig~\ref{fig:chi}-(\textbf{i-l}). In contrast to the above described optimal cloning attack, Eve directly measures the received photon information, where she randomly performs her measurement in bases uniformly chosen at random (intercept). She then prepares a photon in the observed state and transmits it to Bob (resend). When considering the intercept-resend over all MUBs, the fidelity of the states at the output of the channel is given by $F_\mathrm{ir} = 2/(1+d)$. We experimentally perform the intercept-resend attack in our quantum channel by means of two SLMs, i.e. SLM-E1 and SLM-E2, with a SMF in between, see Fig.~\ref{fig:setup}-\textbf{c}. The first SLM (SLM-E1) along with the SMF acts as the intercept section, while the second SLM (SLM-E2) acts as the resend. In the universal case, Eve randomly chooses $\alpha = 1,...,d+1$ and $m=1,...,d$ and simultaneously displays phase elements equivalent to $\langle \psi_m^{(\alpha)}|$ and $|\psi_m^{(\alpha)} \rangle$ on the SLM-E1 and SLM-E2, respectively.

\section{Secret key rates}
{So far, we have shown the feasibility of performing high-dimensional QPT on several quantum communication channels with our experimental implementation and provided a qualitative illustration of the effect of different eavesdropping strategies. In this section, we use the experimental outcomes obtained when performing QPT as our raw key from which we carry a formal security analysis. In particular, we show that the full characterization enables an enhanced estimation of the information leakage to an adversary eavesdropper. This results in a more accurate and favourable estimation of the achievable secret key rate, $K$, of the channel for a given QKD protocol. Furthermore, we demonstrate that the full characterization of the channel becomes even more advantageous when considering larger dimensions.}

Traditionally, the secret key rate of a channel is calculated by evaluating the error rate, $Q$, from the raw key shared by Alice and Bob. Analytical formulae are derived to obtain the secret key rate as a function of the error rate. For instance, the BB84 protocol~\cite{bennett:84} in dimension $d$ has the following simple analytical formula, 
\begin{equation}
	K^{(d)}(Q) = \log_2 (d) - 2 \, h^{(d)} (Q),
	\label{formula:BB84}
\end{equation}
where $h^{(d)}(x):=-x \log_2 (x/(d-1)) - (1-x) \log_2 (1-x)$ is the $d$-dimensional Shannon entropy. This protocol does not require a full characterization of the quantum channel and has a sifting efficiency of 1/2. An extension of the $d$-dimensional BB84 protocol to tomographically complete measurements, yielding a full characterization of the channel, is known as the \emph{six-state} protocol for the specific case of $d=2$. For dimensions that are powers of prime numbers, all $d+1$ MUBs are adopted. The sifting efficiency of this protocol is given by $1/(d+1)$. Nevertheless, in the infinite key limit, the sifting efficiency of the BB84 and the six-state protocol can approach 1, by making the basis choice extremely asymmetric~\cite{lo:05}. This argument also holds for higher dimensions. The secret key rate of this $(d+1)$-MUB protocol as a function of the error rate is given by,
\begin{eqnarray}
	K^{(d)}(Q)&=&\log_2(d) - h^{(d)}\left(\frac{d+1}{d} Q\right) \nonumber \\ && -\frac{d+1}{d} Q \log_2 (d+1).
\label{formula}
\end{eqnarray}

Nonetheless, using such analytical formulae has a number of drawbacks. First, it only considers sifted data and assumes perfect unbiasedness among different MUBs. This assumption may increase the resulting secret key rate, but does not consist of an adequate representation of the channel and may lead to an underestimation of Eve's leaked information. Moreover, the secret key rate is obtained from a coarse-graining of the individual error rates per measurement settings. Thus, given this single parameter, a pessimistic estimation of the secret key rate is obtained due to the limited knowledge of the channel. {Instead, we propose a protocol based on QPT where all the measurement outcomes, both from matching and non-matching bases, are considered in the secret key rate analysis. This protocol based on full characterization of the channel outperforms the counterpart coarse-grained protocols particularly for the case of asymmetric errors, which is often the case in experiments.

\begin{figure}[ht!]
	\begin{center}
	\includegraphics[width=1\columnwidth]{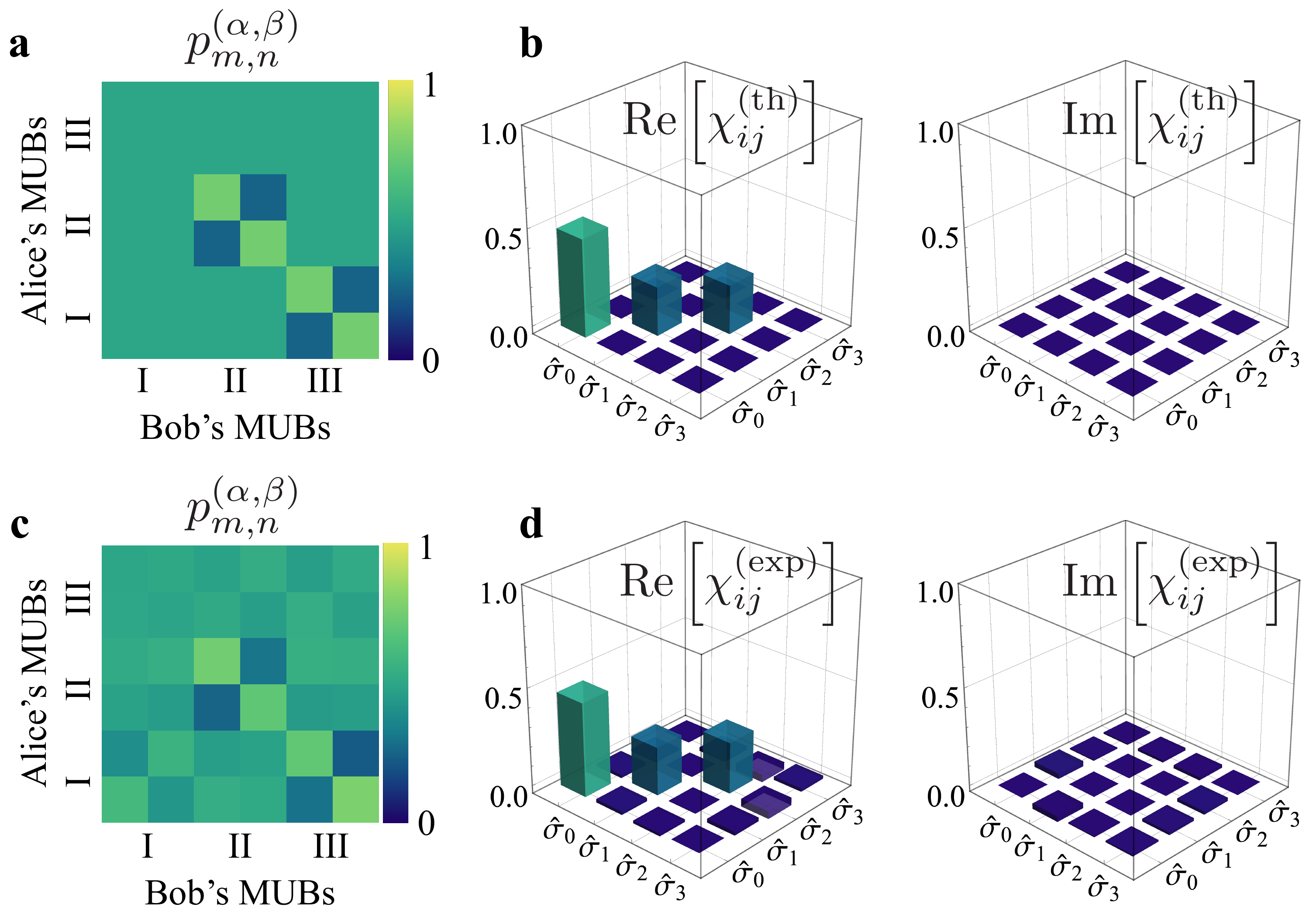}
	\caption[]{{\textbf{Asymmetric intercept-resend.} \textbf{a} Theoretical and \textbf{c} experimental probability of detection matrices, as given in Eq.~\ref{probability2}, corresponding to the asymmetric process of an intercept-resend eavesdropping strategy over MUB II and III in dimension 2. \textbf{b} Theoretical and \textbf{d} experimental real and imaginary part of the process matrix.}}
	\label{fig:asymchi}
	\end{center}
\end{figure}

In our analysis, the secret key rate of a channel is estimated by considering the Devetak-Winter formula~\cite{devetak:05}.} To do so, we recast our prepare-and-measure QKD protocol into an entanglement-based scheme using the source-replacement scheme~\cite{scarani:09}, where Alice and Bob share pairs of entangled photons represented by the density matrix $\rho_{AB}$. In general, the secret-key rate is obtained according to,
\begin{equation}
	K =  \min_{\rho_{AB} \, \in \, {\cal C}} \left[ H\left( Z_A | E \right) - H\left( Z_A | Z_B \right) \right],
\end{equation}
where ${\cal C}$ is the set of physical density matrices that is consistent with the experimental constraints, e.g. error rates {or measurement outcomes}; $H\left( X| Y \right) := H \left( \rho_{XY} \right) - H \left( \rho_Y \right)$ is the conditional von Neumann entropy, with $H \left( \rho \right) := - \mathrm{Tr} \left[ \rho \log_2 \rho  \right]$; $Z_A$ and $Z_B$ are the sets of positive operator valued measures (POVM) associated with Alice's and Bob's measurement settings, respectively. Finally, $\rho_{Z_A Z_B}$ and $\rho_{Z_A E}$ are given by,
\begin{equation}
	\rho_{Z_A Z_B} = \sum_{j,k} \mathrm{Tr} \left[ \left( Z_A^j \otimes Z_B^k \right) \rho_{AB} \right] | j \rangle \langle j | \otimes | k \rangle \langle k |,
\end{equation}
\begin{equation}
	\rho_{Z_A E} = \sum_{j} | j \rangle \langle j | \otimes \mathrm{Tr}_A \left[ \left( Z_A^j \otimes \hat{I}  \right) \rho_{AE} \right],
\end{equation}

where $\rho_{ABE}$ is the tripartite density matrix shared by Alice, Bob and Eve, respectively. Hence, every additional experimental measurement is included in the minimization, which results in a greater, or equal, secret key rate. In the limiting case where we have a full characterization of the channel, $\rho_{AB}$ is fully reconstructed and is given by the Choi matrix, $\rho_{\cal E}$, mentioned earlier. Thus the secret key rate may be directly calculated from $\rho_{AB}$,
\begin{equation}
	K =  H\left( Z_A | E \right) - H\left( Z_A | Z_B \right),
\label{DW}
\end{equation}
where all experimental measurements are taken into consideration, even for the case of mismatching MUBs. 

\begin{figure*}[ht!]
	\begin{center}
	\includegraphics[width=1.75\columnwidth]{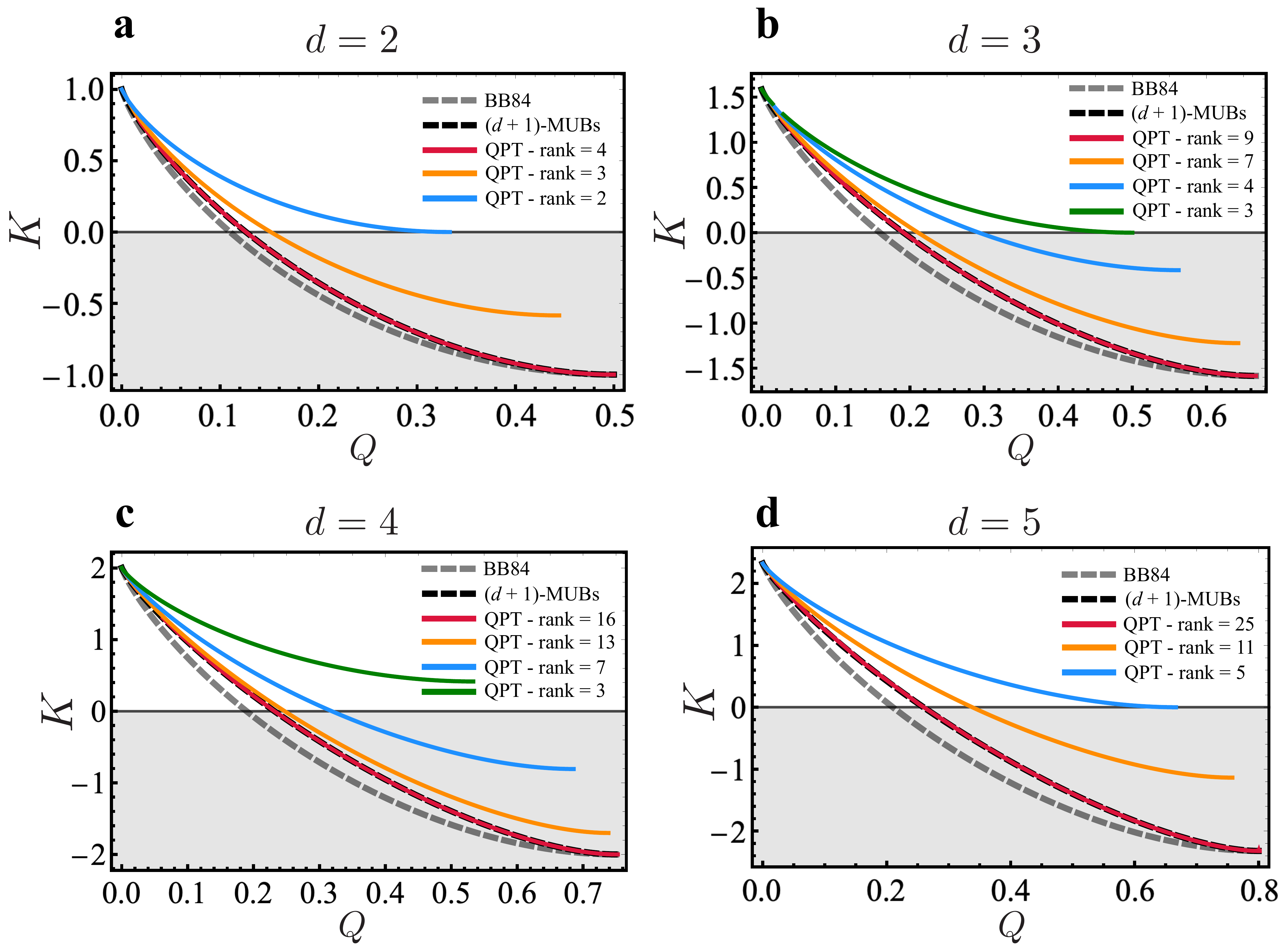}
	\caption[]{{\textbf{Security analysis for noise models with varying levels of symmetry.} \textbf{a}-\textbf{d} The secret key rate is calculated for different scenarios in dimensions ranging from 2 to 5. The secret key rate for the BB84 and the $(d+1)$-MUBs protocols are directly calculated from Eq.~\ref{formula:BB84} and Eq.~\ref{formula}, respectively, where the level of noise is characterized using the error rate. Using the process matrix representation, we then consider noisy channels corresponding to diagonal trace-preserving process matrices of different ranks. The secret key rate is directly calculated using Eq.~\ref{DW}, and an error rate is determined for comparison with the coarse-grained protocols. For the case of a process matrix of rank $r=d^2$ with constant diagonal elements, i.e. Eq.~\ref{eq:chisym}, we retrieve the case of completely symmetric noise, i.e. Eq.~\ref{formula}. Asymmetric noise is then simulated by considering process matrices of lower ranks.} }
	\label{fig:noise}
	\end{center}
\end{figure*}

{In the prepare-and-measure scheme, we can now determine the set of POVMs corresponding to Alice and Bob's generation and measurement apparatuses. At Alice's preparation stage, we have $ Z_A = \{ | 0 \rangle \langle 0 | , ... , | d-1 \rangle \langle d-1 |  \}$, since Alice has the freedom to generate any state at random. However, at Bob's measuring stage, we consider two different sets of POVMs, i.e. $Z_B^\mathrm{sort} = \{ | 0 \rangle \langle 0 | , ... , | d-1 \rangle \langle d-1 |   \}$ and $Z_B^\mathrm{filter} = \{ | i \rangle \langle i |\}$, where the former corresponds to a sorting-type measurement and the latter corresponds to a filter-type measurement, and Alice and Bob post-select on the event where Bob successfully obtains a measurement of the state $|i\rangle \langle i |$ with $i \in \{0,...,d-1\}$. When carrying out the security analysis, one can see that the secret key rate, $K$, for the cases of $Z_B^\mathrm{sort}$ and $Z_B^\mathrm{filt}$ only differs from an overall efficiency of $1/d$. Thus, we carry out the high-dimensional QKD protocols based on QPT with a filter-type measurement scheme at Bob's stage using an SLM. There exists efficient sorters for the OAM computational and discrete Fourier transform basis~\cite{berkhout:10}; however, the establishment of arbitrary multi-outcome measurement devices remains an open challenge for quantum applications~\cite{fontaine:18}.

For the full characterization of our high-dimensional quantum channel via QPT with no explicit eavesdropping strategy applied, we calculate the corresponding secret key rate and compare it to a coarse-grained estimation of the channel, see Fig.~\ref{fig:keyrate} \textbf{a}. In the case of explicit eavesdropping strategies, such as optimal cloning and intercept-resend, the secret key rate will obviously be negative, see Fig.~\ref{fig:keyrate} \textbf{b}-\textbf{c}, and no secret key can be exchanged between Alice and Bob. We observe that for larger dimensions, the full characterization secret key rate performs better compared to the achievable secret key rate obtained from a coarse-graining of the channel. This may be due to the fact that, as one considers higher-dimensional states, the symmetry assumption involved in coarse-graining is increasingly not fulfilled. Therefore, full characterization of quantum channels may offer a means by which the full potential of high-dimensional protocols is exploited. In highly asymmetric scenarios, which is often the case due to systematic errors such as misalignments, full characterization may even surpass the performance of coarse-grained protocols, such as the BB84 protocol, when considering the lower sifting efficiency of the full characterization protocol in the finite-key scenario.

Finally, in order to better understand the advantage of full characterization protocols compared to their coarse-grained counterparts, we carry out a security analysis for different noise models. In particular, we take advantage of the process matrix representation of the quantum channel to simulate different levels of symmetry in noise. In the limit, where the noise is completely symmetric over all states of all bases, the process matrix of the channel can be described as following,

\begin{eqnarray}
\chi_{ij}^{r=d^2} &=&  \frac{F(d+1)-1}{d}  \delta_{0i} \delta_{0j} \nonumber \\ && + \frac{(1-F)}{2 (d-1)} \left(  \delta_{ij} -\delta_{0i} \delta_{0j} \right),
\label{eq:chisym}
\end{eqnarray}
where $\chi_{ij}^{r=d^2}$ is a $d^2 \times d^2$ diagonal matrix with a rank of $r=d^2$, and $F$ determines the level of noise in the channel. For instance, the result of such a process on an input state, say $\rho_\mathrm{in}=|0\rangle \langle 0|$, is given by the output state $\rho_\mathrm{out}=F|0\rangle \langle 0| + (1-F)/(d-1) \sum_{i=1}^{d-1} |i\rangle \langle i|$. The processes of universal optimal cloning and universal intercept-resend fall under this category of process matrices, as can be seen in Fig.~\ref{fig:chi}, with fidelities $F_\mathrm{cl}=1/2+1/(1+d)$ and $F_\mathrm{ir} = 2/(1+d)$, respectively. Here, we consider another type of eavesdropping strategy that is not considered universal. We experimentally perform QPT of an intercept-resend attack where the eavesdropper only consider 2 out of the 3 MUBs available to Alice and Bob in dimension 2, see Fig.~\ref{fig:asymchi}. This would be an appropriate attack in the case of BB84, where only two MUBs are utilized by Alice and Bob. However, in this case the noise added by the process to the output state is no longer symmetric. We observe state- and basis-dependent state fidelities which can be understood as asymmetric noise. Interestingly, the process matrix of the 2-MUB intercept-resend attack, see Fig.~\ref{fig:asymchi}, has only 3 non-zero diagonal elements compared to 4 non-zero diagonal elements for the case of universal attacks, see Fig.~\ref{fig:chi}, in dimension 2. Inspired, by the shape of the process matrix in this simple case, we investigate different asymmetric noise models by varying the rank of the process matrix

For a process matrix of rank $r=d^2$, we retrieve the secret key rate formula for the coarse-grained $(d+1)$-MUBs protocol, i.e. Eq.~\ref{formula}, as can be seen in Fig.~\ref{fig:noise}. In dimension 2, we may then consider channels with process matrices of other ranks, corresponding to different degrees of noise symmetry. In particular, we consider trace-preserving process matrices of rank 3, i.e. ${\chi_{ij}^{d=2,r=3}= F \delta_{0i} \delta_{0j} + (1-F)/2 (\delta_{1i} \delta_{1j} + \delta_{2i} \delta_{2j})}$, and rank 2, i.e. ${\chi_{ij}^{d=2,r=2}= F \delta_{0i} \delta_{0j} + (1-F)(\delta_{1i}\delta_{1j})}$. As can be seen in Fig.~\ref{fig:noise}-\textbf{a}, noise corresponding to a process matrix of rank 2 or 3, leads to a significant improvement in secret key rate compared to a coarse-grained characterization of the channel using the error rate. Interestingly, for the case of the rank 2 process matrix of the form given above, i.e. ${\chi_{ij}^{r=2}}$, although the resulting error rate is well above the traditional error threshold, the secret key rate remains positive all the way up to $Q=0.33$. We consider trace-preserving process matrices of rank 7, 4, and 3, in dimension 3, rank 13, 7, and 3, in dimension 4, and rank 11 and 5, in dimension 5, i.e. \newline

\begin{eqnarray}
\chi_{ij}^{d=3,r=7}&=& F \delta_{0i} \delta_{0j} + \frac{1}{4}(1-F) \sum_{k=1}^6 \delta_{ki} \delta_{kj} \nonumber \\
\chi_{ij}^{d=3,r=4}&=& F \delta_{0i} \delta_{0j} + \frac{1}{2}(1-F) \sum_{k=1}^3 \delta_{ki} \delta_{kj} \nonumber\\
\chi_{ij}^{d=3,r=3}&=& F \delta_{0i} \delta_{0j} + \frac{3}{4}(1-F) \, (\delta_{1i} \delta_{1j} + \delta_{8i} \delta_{8j})\nonumber \\
\chi_{ij}^{d=4,r=13}&=& F \delta_{0i} \delta_{0j} + \frac{1}{16}(1-F) \sum_{k=1}^{12} \delta_{ki} \delta_{kj}\nonumber\\
\chi_{ij}^{d=4,r=7}&=& F \delta_{0i} \delta_{0j} + \frac{1}{3}(1-F) \sum_{k=1}^6 \delta_{ki} \delta_{kj}\nonumber\\
\chi_{ij}^{d=4,r=3}&=& F \delta_{0i} \delta_{0j} + (1-F) ( \delta_{1i} \delta_{1j}+\delta_{12i} \delta_{12j} )\nonumber\\
\chi_{ij}^{d=5,r=11}&=& F \delta_{0i} \delta_{0j} + \frac{1}{4}(1-F) \sum_{k=1}^{10} \delta_{ki} \delta_{kj}\nonumber\\
\chi_{ij}^{d=5,r=5}&=& F \delta_{0i} \delta_{0j} + \frac{1}{4}(1-F) \sum_{k=21}^{24} \delta_{ki} \delta_{kj},
\end{eqnarray}
respectively. In all cases, we can observe a clear advantage of using a full characterization scheme compared to the error rate channel assessment. Interestingly, for all dimensions we find cases where after the transmission, the coarse-grained errors would label the channel as insecure, while our more detailed analysis still enables a secure communication. This advantage is possible because our analysis allows to distinguish between cases where an eavesdropper has not gained enough information about the state although they have transformed it such that Alice and Bob will have large error rates. 
}

\section{Conclusion}

In summary, full characterization of quantum processes via QPT is an invaluable tool for high-dimensional quantum information processing. In particular, the complexity resulting from the generation and detection of the high-dimensional states involved may be fully characterized. In the study of quantum channels for quantum communications, full characterization via QPT turns out to be a beneficial resource that allows one to take full advantage of the potential high-dimensional nature of the protocols at play to increase the overall secret key rate.

\noindent
\vspace{0.5 EM}

\noindent\textbf{Acknowledgments}
This work was supported by Canada Research Chair (CRC); Canada Foundation for Innovation (CFI); Canada Excellence Research Chairs, Government of Canada (CERC); Canada First Research Excellence Fund (CFREF); Natural Sciences and Engineering Research Council of Canada (NSERC). D.K. would like to acknowledge the project no IGA-PrF-2018-003.

\onecolumn\newpage

\end{document}